\documentclass[12pt,twoside]{article}   
\usepackage[super,sort,comma]{natbib}
\usepackage{caption}
\usepackage{subcaption}

\usepackage{fancyhdr}		

\usepackage[section]{placeins}   %

\usepackage{graphicx}

\makeatletter \renewcommand\@biblabel[1]{$^{#1}$} \makeatother
\newcommand{\cen}[1]{\begin{center} #1 \end{center}}

\usepackage[mathlines]{lineno}

\usepackage{chemformula}

\usepackage{upgreek}

\usepackage{hyperref}
\hypersetup{ colorlinks,
    citecolor=blue,
    filecolor=blue,
    linkcolor=blue,
    urlcolor=blue
}

\usepackage{xcolor}

\definecolor{gray}{rgb}{0.6,0.6,0.6}
\definecolor{red}{rgb}{0.85,0,0}
\definecolor{green}{rgb}{0,0.85,0}
\definecolor{blue}{rgb}{0,0,0.85}
\definecolor{beige}{rgb}{0.92,0.87,0.78}
\usepackage[all]{hypcap}    

\begin{document}

\cen{\sf {\Large {\bfseries Does FLASH deplete Oxygen?  Experimental Evaluation for Photons, Protons and Carbon Ions. } \\  
\vspace*{10mm}
	Jeannette Jansen$^{1,2}$, Jan Knoll$^{1,2}$,  Elke Beyreuther$^{3,4}$, Jörg Pawelke$^{3,5}$, Raphael Skuza$^{1,2}$, Rachel Hanley$^{1,2}$, Stephan Brons$^{6}$ , Francesca Pagliari$^{1}$, Joao Seco$^{1,2}$} \\
$^{1} $ Division of Biomedical Physics in Radiation Oncology, German Cancer Research Center (DKFZ), Heidelberg, Germany\\
$^{2} $ Faculty of Physics and Astronomy, Ruprecht-Karls-University Heidelberg, Germany\\
$^{3} $ OncoRay – National Center for Radiation Research in Oncology, Faculty of Medicine and University, Hospital Carl Gustav Carus, Technische Universität Dresden, Helmholtz-Zentrum Dresden – Rossendorf, Dresden, Germany\\
$^{4}$ Helmholtz-Zentrum Dresden – Rossendorf (HZDR), Institute of Radiation Physics, Dresden, Germany\\
$^{5}$ Helmholtz-Zentrum Dresden – Rossendorf (HZDR), Institute of Radiooncology - OncoRay, Dresden, Germany \\
$^{6}$ Heidelberg Ion-Beam Therapy Center (HIT), Heidelberg, Germany
\vspace{5mm}\\
Version typeset \today\\
}

\pagenumbering{roman}
\setcounter{page}{1}
\pagestyle{plain}
Author to whom correspondence should be addressed: Joao Seco. email: j.seco@dkfz-heidelberg.de\\

\begin{abstract}
\noindent {\bf Purpose:} To investigate experimentally, if FLASH irradiation depletes oxygen within water for different radiation types such as photons, protons and carbon ions.\\
{\bf Methods:} This study presents measurements of the oxygen consumption in sealed, 3D printed water phantoms during irradiation with X-rays, protons and carbon ions at varying dose rates up to 340\,Gy/s. The oxygen measurement was performed using an optical sensor allowing for non-invasive measurements.\\
{\bf Results:} Oxygen consumption in water only depends on dose, dose rate and linear energy transfer (LET) of the irradiation. The total amount of oxygen depleted per 10\,Gy was found to be 0.04\,\% atm - 0.18\,\% atm for 225\,kV photons, 0.04\,\% atm - 0.25\,\% atm for 224 MeV protons and 0.09\,\% atm - 0.17\,\% atm for carbon ions.
Consumption depends on dose rate by an inverse power law and saturates for higher dose rates because of self-interactions of radicals. Higher dose rates yield lower oxygen consumption. No total depletion of oxygen was found for clinical doses.\\
{\bf Conclusions:} FLASH irradiation does consume oxygen, but not enough to deplete all the oxygen present. For higher dose rates, less oxygen was consumed than at standard radiotherapy dose rates. No total depletion was found for any of the analyzed radiation types for 10\,Gy dose delivery using FLASH. 
\end{abstract}

\newpage     

The table of contents is for drafting and refereeing purposes only. Note
that all links to references, tables and figures can be clicked on and
returned to calling point using cmd[ on a Mac using Preview or some
equivalent on PCs (see View - go to on whatever reader).
\tableofcontents

\newpage

\setlength{\baselineskip}{0.7cm}      

\pagenumbering{arabic}
\setcounter{page}{1}
\pagestyle{fancy}
\section{Introduction}

Over the last years, research on the irradiation with high dose rates (i.e. FLASH irradiation) became increasingly important. \textit{In vivo}, studies showed a radio-protective effect in healthy tissue when irradiated with electrons at high dose rates ($>$ 40\,Gy/s) whereas the tumor control probability remained comparable to usual (clinical) dose rates of around 2\,Gy/min\cite{Favaudon2014}.
Applied to a clinical setup, FLASH dose rate irradiation could therefore enlarge the therapeutic window, i.e. healthy tissue is protected and irradiation with higher doses in the tumor is made possible. \\
Although this differential effect of radio-protection of the normal tissue has been studied and confirmed already \textit{in vivo} \cite{vozenin2019advantage}, the mechanism behind the FLASH effect still remains unknown\cite{boscolo2020impact}.
It is believed that dissolved oxygen in the cellular cytoplasm plays a major role:
Early findings in the 1960s and 1970s showed a hypoxic-like cell survival behavior when Escherichia coli bacteria were irradiated with ultra-high dose rates of X-rays\cite{Weiss1974}.
Oxygen measurements for the same experimental setup showed a decrease of oxygen during irradiation. Similar results were obtained in HeLa cells\cite{town1967effect} and chinese hamster cells\cite{berry1969survival}. One of the possible mechanisms to explain the FLASH effect nowadays is that oxygen is depleted during irradiation which causes a hypoxic environment in the irradiated volume.
Hypoxic tissues are known to be 2-3 times more radio-resistant than normoxic tissues \cite{tinganelli2015kill}.   
As tumors are mostly hypoxic (i.e. the \ch{O2} concentration is lower than 0.5\,\% atm) and healthy tissue (with an \ch{O2} concentration of 1\,\% - 11\,\% \cite{carreau2011partial}) is not, the oxygen depletion theory is of current interest as it could explain the observed radio-protective effect. \\
In this case, the consumption of oxygen, which can lead to complete depletion, is due to radiolysis. This process creates various radicals which then react with oxygen and result in a decrease in molecular oxygen.\\

\subsection{Radiolysis}
Ionizing radiation causes radiolysis of water molecules producing a range of reactive species (see Table \ref{tab:G}). On short timescales after irradiation (until $10^{-12}$\,s), a high production of \ch{e_{aq}^-} and \ch{H^.} is observable. In presence of molecular oxygen dissolved in water, these species can interact with \ch{O2} which leads to the production of superoxide (\ch{O_2^.}) and \ch{HO2^.}. These products react further with each other on time scales until up to $10^{-6}$\,s. After $10^{-6}$ seconds, the production of most radicals is in a stable regime and will not cause further reactions \cite{boscolo2020impact}. For the studies presented, those reactions, which have \ch{O2} as direct product or educt, are of main interest, i.e. reactions in which \ch{O2} is directly consumed or produced (see Table \ref{tab:Ox_consumption} + \ref{tab:Ox_production}). \\

\begin{table}[h!]
	\centering
	\begin{tabular}{c|c|c|ccc}
		\hline
		            & Photons (1.2\,MeV) &    Protons    &               &    $^{12}$C  &                \\
		G value of: &         & 1\,keV/$\mu$m & 10\,keV/$\mu$m & 15\,keV/$\mu$m & 20\,keV/$\mu$m \\ \hline
		 \ch{^.OH}  &   2.7   &      2.5      &      1.7      &      1.5       &      1.3       \\
		 \ch{e_{aq}^-}   &   2.6   &      2.3      &      1.5      &      1.3       &      1.1       \\
		 \ch{H2O2}  &  0.70   &     0.68      &     0.83      &      0.87      &      0.89      \\
		 \ch{H^.}   &  0.60   &     0.63      &     0.69      &      0.67      &      0.65      \\
		  \ch{H2}   &  0.45   &     0.49      &     0.62      &      0.68      &      0.71      \\ \hline
	\end{tabular}
	\caption{G-values in molecules/100\,eV of primary radicals for photon radiation\cite{joseph2008combined} and for ion radiation\cite{meesungnoen2005high} in water at 25\,$^\circ$C and pH 7.}\label{tab:G}
\end{table}

\vspace{1cm}

\begin{table}[h!]
	\centering
\begin{tabular}{llr}
	\hline
	Reaction && Reaction rate $k$ $\left[10^{10} \mathrm{dm}^3\mathrm{mol}^{-1}\mathrm{s}^{-1}  \right]$ \\
	\hline
	\ch{e_{aq}^- + O2}&\ch{ -> O2^{.-}}  & 1.9 \\
	\ch{H^. + O2}&\ch{ -> HO_2^.}& 2.0 \\
	\hline
\end{tabular}
\caption{Reactions consuming oxygen}\label{tab:Ox_consumption}   
\end{table}

\vspace{1cm}

\begin{table}[h!]
	\centering
\begin{tabular}{llr}
	\hline
	Reaction && Reaction rate $k$ $\left[10^{10} \mathrm{dm}^3\mathrm{mol}^{-1}\mathrm{s}^{-1}  \right]$ \\
	\hline
	\ch{OH^. + HO_2^.}& \ch{ -> O2 + H2O} &1.0\\
	\ch{OH^. + O2^{.-}}& \ch{ -> O2 + OH^-} &0.9\\
	\ch{HO_2^. + HO2^.}& \ch{ -> H2O2 + O2} &0.000076\\
	\ch{HO_2^. + O2^{.-}}& \ch{ -> O2 + HO2^-} &0.0085\\
	\hline
\end{tabular}
\caption{Reactions producing oxygen  
}\label{tab:Ox_production}   
\end{table}

For deeper understanding, it is crucial to quantify the amount of oxygen that is consumed as a result of irradiation. This amount is given by G-values, describing the amount of molecules produced per 100\,eV imparted in the water (Table \ref{tab:G}).

Various simulation studies have been published to date (\cite{boscolo2020impact}, \cite{pratx2019computational}, \cite{petersson2020quantitative}), but there is a lack of measurement data and systems that not only measure oxygen consumption before and after irradiation, but can also be used \textit{in-vitro} (i.e. together with cell culture) and online.
The aim of this work is therefore to investigate experimentally the oxygen consumption in pure water as a potential mechanism of FLASH using an online oxygen meter. 
Thereby, the study is designed to cover a broad range of radiation types (X-ray, p and $^{12}$C radiation) and dose rates (2 Gy/min - 340 Gy/s).

\section{Materials and Methods}
\subsection{Preparation of Oxygen Meter for Measuring Dissolved Oxygen}

For the experimental part of the study, radiolysis of water and the resulting oxygen consumption were investigated using the solid optical sensors TROXSP5 from PyroScience GmbH in a 3D-printed water phantom.

\begin{figure}[h!]
	\centering
	\includegraphics[width = .6\linewidth]{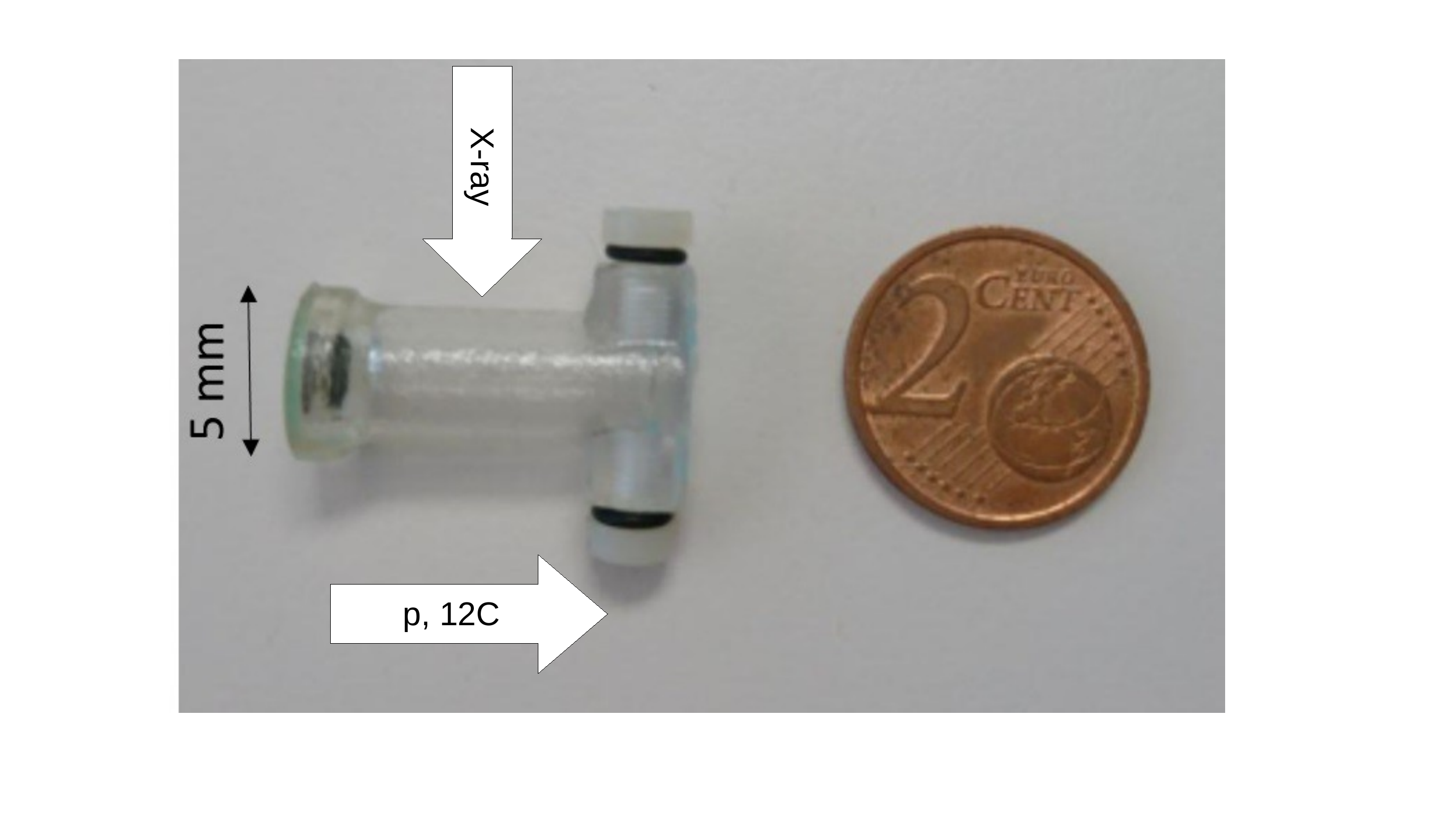}
	\caption{3D-printed water phantom of 5\,mm inner diameter. \ch{O2} sensor (black) is placed on the inside on the left end, facing away from the beam. At this side, the optical fiber can be coupled to the phantom. On the right, facing the beam, two openings for filling the phantom are visible, which can be closed with plastic screws. O-rings were used between the screws and the phantom for additional air tightness. The white arrows show the beam's direction for the respective beam types.}
	\label{fig:phantom}
\end{figure}

The water phantoms suitable for this study had to be airtight, preparable in different geometries to adapt to different irradiation beam set ups and transparent to read the sensor optically.
To fulfill these requirements, phantoms were 3D-printed out of the material VeroClear (Stratasys Ltd., Israel), a rigid, colourless and transparent material. The optical sensor was glued with silicone into the phantom and the phantom was filled with de-ionized water. 
The oxygen dissolved in water was measured via a fluorescent layer in the sensor, which was read-out with the purchased system FireStingO2 (FSO2-4, PyroScience GmbH). With the FireStingO2, the sensor is excited at 650\,nm wavelength and emits light in the near infrared regime. This signal is then further processed in the FireStingO2. Time resolution of around 400\,ms can be achieved.\\
Since the read-out is performed optically, it was possible to measure the changes in oxygen concentration inside a water phantom non-invasively. This is the main advantage over studies with other commonly used oxygen probes where measurements are usually an invasive process.\\
The phantom itself was 20\,mm long, of cylindrical shape and produced for multiple beam diameters, to achieve a uniform dose distribution within the phantom. Hence, the phantom's diameter was constructed significantly smaller than the FWHM of the respective beam used for irradiation. For the present study, phantoms of 5\,mm (Fig. \ref{fig:phantom}) in diameter were applied for photon radiation, and phantoms of  2\,mm in diameter were applied for proton and $^{12}$C radiation.

\subsection{Photons, Protons and Carbon Ion Beams}	
To investigate the oxygen consumption as a function of dose and dose rate (i.e. the amount of dose deposited per time interval), the water phantom with the sensor was irradiated at different dose rates with vertical beams of 225 kV photons (Faxitron MultiRad225, Faxitron Bioptics, LLC). Irradiations with carbon ions were performed at Heidelberg Ion beam Therapy facility HIT, Germany at up to 9.5\,Gy/s peak dose rate using the horizontal beam line in the QS room of HIT. Irradiations with protons were performed at OncoRay, Dresden, Germany at dose rates up to 340\,Gy/s using the horizontal beam line in the experimental room of the University Proton Therapy facility. The applied beam parameters can be found in Table \ref{tab:beam_params}. For both proton and carbon ion setups, the phantom was irradiated with high energy particles, i.e. in the plateau region of the depth-dose-curve of the beam.

\vspace{1cm}
\begin{table}[h!]
	\begin{tabular}{|l|l|l|l|l|l|}
		\hline
		& Energy     & LET$_{\ch{H2O}}${[}keV/$\mu$m{]}   &Av. DR {[}Gy/s{]} & Spill DR {[}Gy/s{]} & Beam Structure \\ \hline
		X-ray         & 225 kV     &$\sim$ 1.7                         & 0.03  - 52          & -& continuous \\ \hline
		p                 & 224\,MeV/u & 0.42                   & 0.03 - 340           & -& 2\,ns beam on \\
 & & & & &+ 8\,ns beam off \\
 \hline
		$^{12}\mathrm{C}$ & 400\,MeV/u & 10.89                     & 0.06 - 2.4            & 0.12 - 5.1 & 1.5\,s - 4.9\,s on\\
 & & & & & + 4\,s - 5\,s off\\
 \hline
$^{12}\mathrm{C}$		& 150\,MeV/u & 19.47                      & 0.04 - 1.8         & 0.06 - 9.5& 1.5\,s - 4.9\,s on\\
 & & & & & + 4\,s - 5\,s off\\
 \hline
	\end{tabular}
	\caption{Beam parameters used for irradiation. Linear energy transfer (LET) was calculated using ICRU49\cite{ICRU49} and ICRU73\cite{ICRU73}. Photon LET was estimated\cite{LANNUNZIATA20031}.  }\label{tab:beam_params}
\end{table}

\subsection{Measurement Setup}
\begin{figure}[h!]
	\centering
	\includegraphics[width = .5\linewidth]{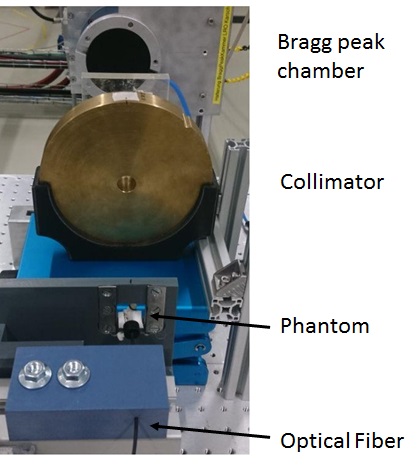}
	\caption{Experimental set up at OncoRay, Dresden for irradiation with 224\,MeV protons. The beam passed a Bragg peak chamber (model T34070-2.5 from PTW) for dose monitoring and was shaped with a scatterer of 24\,mm PMMA equivalent thickness and a 20\,mm thick brass collimator to deliver a homogenous beam spot. The phantom was held in place with a sample holder.}
	\label{fig:Oncoray}
\end{figure}

For measuring oxygen consumption, the phantom was filled entirely with pure, double-deionized water, without air bubbles, and closed with plastic screws. The oxygen content of the water was changed using a  Sci-Tive hypoxic chamber (Baker Ruskinn), in which \ch{N2} is used as air substitute. The pure water was placed in the hypoxic chamber for 2 days until the desired \ch{O2} level was reached and the phantoms were then filled with hypoxic water inside the chamber. Then, the phantom was placed at the beam's central beam axis. For proton and carbon ion measurements, the phantom's cylinder axis was placed on the beam axis. For photon measurements, the phantom was placed perpendicular to the beam instead. By that, the phantoms were always positioned horizontally to ensure homogeneous water composition. On the sensor's side of the phantom, a fiber holder was placed to connect the phantom's sensor to the FireStingO2 meter with a 2\,m long optical fiber of 2.2\,mm diameter. The FireStingO2 meter was then connected to the laptop for data acquisition. The obtained values for \ch{O2} concentration are in a range of 0\,\% - 20.95\,\% (air saturated).
Figure \ref{fig:Oncoray} shows a photograph of the experimental setup for proton radiation. 

\subsection{Beam Microstructure}
For defining the dose rate, recent studies have raised the importance to distinguish continuous wave (CW) and spilled (pulsed) beams\cite{wilson2020}. In order to achieve the same average dose rate in a spilled beam compared to a CW beam, a much higher dose rate would be required in each pulse of a spilled beam to compensate for the beam pauses.
Hence, for spilled beams, it is crucial to take both the pulse dose rate (i.e. the dose rate obtained in one spill) and the average dose rate into account \cite{esplen2020physics}.

In this study, clear CW structure was obtained in X-rays. Protons at OncoRay show a quasi-continuous structure: 2\,ns beam-on time is followed by 8\,ns beam-off. At HIT, the beam shows a spill-like structure: A continuous beam during spill duration of 1.5-4.9\,s and a time between two spills of around 4-5\,s in which no particles and hence no dose is delivered. The study presented here shows both the oxygen consumption in pulsed beams and in CW beams.\\

\subsection{Dosimetry}
The dosimetry at phantom site for photon irradiation was carried out using a Semiflex ionization chamber (IC, type number TM31010, PTW, Germany). For carbon ion irradiation, a PinPoint chamber (type number TM31015, PTW, Germany) was used and both respectively coupled to a Unidos electrometer (PTW, Germany). Proton dosimetry was achieved using an Advanced Markus IC (type number: 34045, PTW), coupled to an Unidos electrometer as well.  For the maximum dose rate of 340 Gy/s applied, a small saturation correction $k_{sat}$  of 1.01 was determined for the Markus IC \cite{karsch2016derivation}. Therefore, recombination effects can be neglected and no dose rate dependent saturation correction was applied.

\begin{figure}
	\centering
	\begin{subfigure}[b]{0.35\textwidth}
		\centering
		\includegraphics[width = \linewidth]{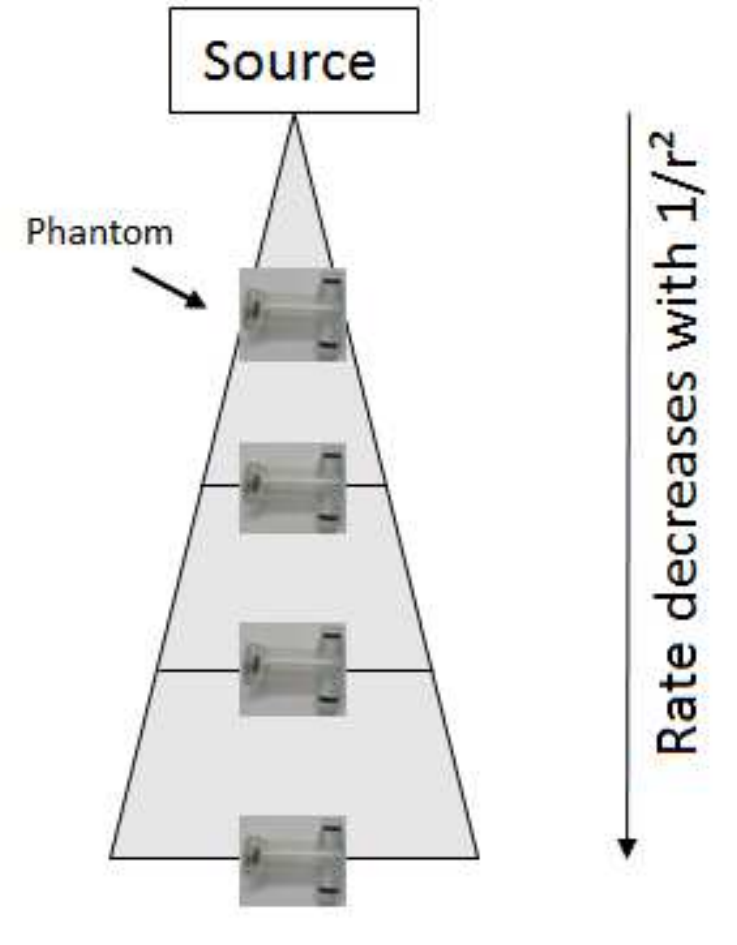} 
		\caption{}
		\label{fig:cone}
	\end{subfigure}
	\hfill
	\begin{subfigure}[b]{0.55\textwidth}
		\centering
		\includegraphics[width = \linewidth]{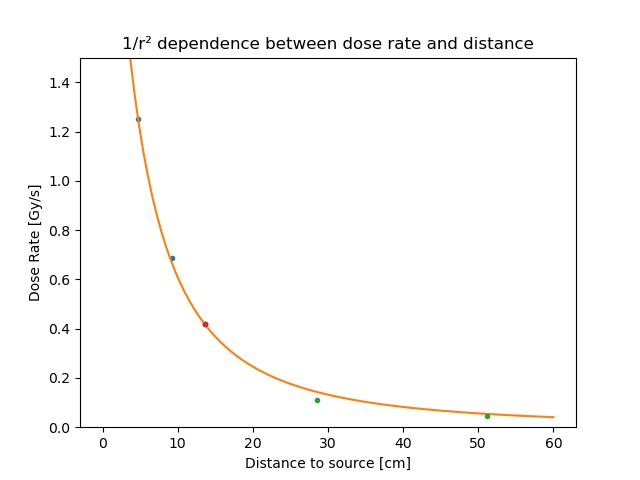} 
		\caption{}
		\label{fig:dose-distance}
	\end{subfigure}
	\caption{(a) Schematic of the conically shaped beam in Faxitron225. (b) Dose rate depends proportionally on $1/r^2$, $r$ being the distance to the beam source. Fit curve in (b) was used to determine dose rates close to source.}
	\label{fig:cone_total}
\end{figure}

In the experiments with photons presented in this study, it was possible to take additional advantage of the beam's geometry: The beam is conically shaped (as schematically shown in Figure \ref{fig:cone}) and the dose is hence inversely proportional to the squared distance from the source, i.e. $D \sim 1/r^2$. Accordingly, the same must apply to the dose rate, i.e. $\dot{D} \sim 1/r^2$. Therefore, irradiating the phantom at different distances from the source automatically leads to a $1/r^2$-dependent dose rate (see Figure \ref{fig:dose-distance}) that can be used for further measurements.
In the experiments with protons and carbon ions, dose rate and dose was changed by setting the beam current and irradiation time in the beam control system.

\section{Results}
\subsection{Oxygen Consumption for Varying Water Phantom Volumes}
\begin{figure}[h!]
	\centering
	\begin{subfigure}[b]{0.45\textwidth}
	\centering
	\includegraphics[width = \linewidth]{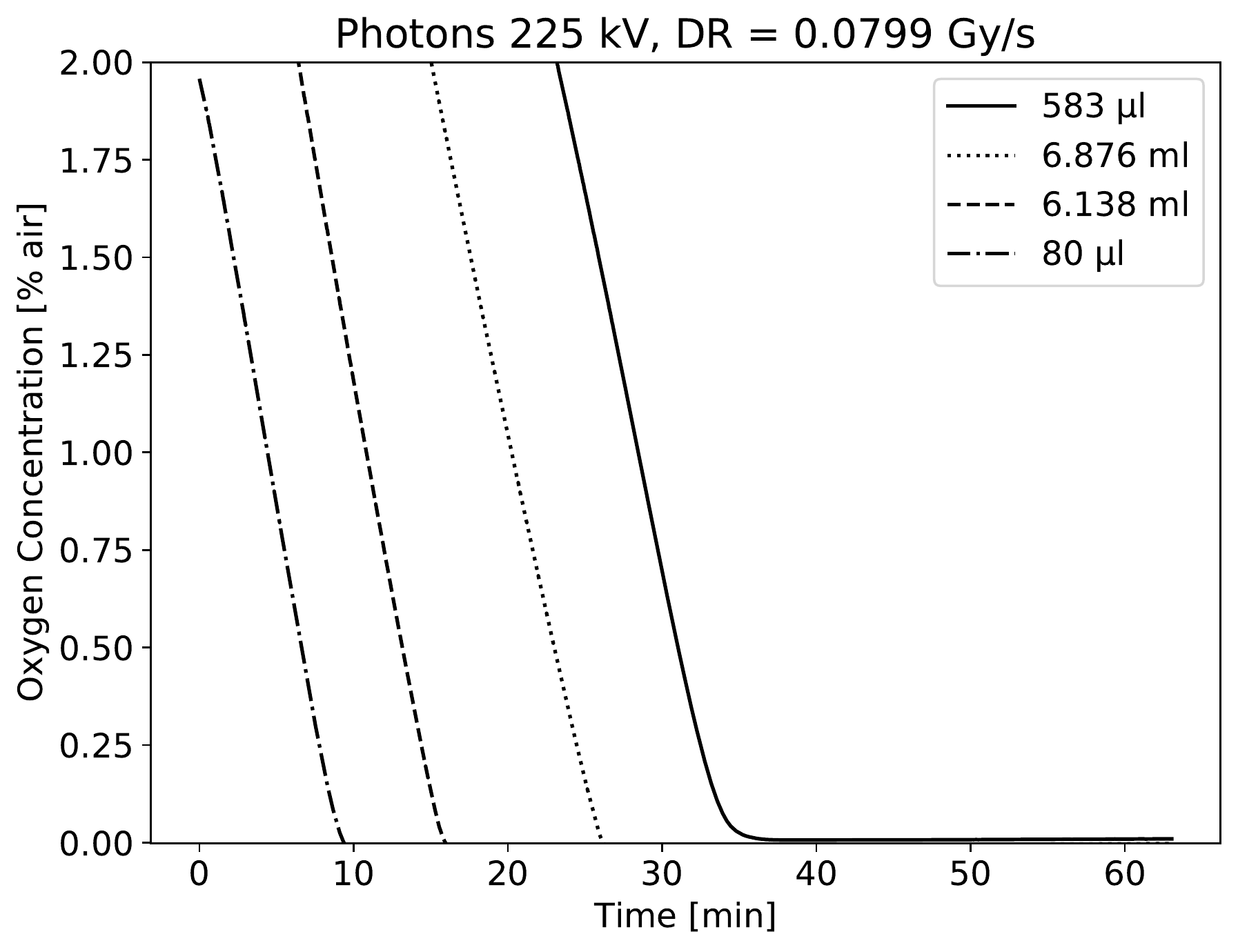}
	\caption{}	\label{fig:volumes}
	\end{subfigure}
	\hfill
	\begin{subfigure}[b]{0.45\textwidth}
	\centering
	\includegraphics[width = \linewidth]{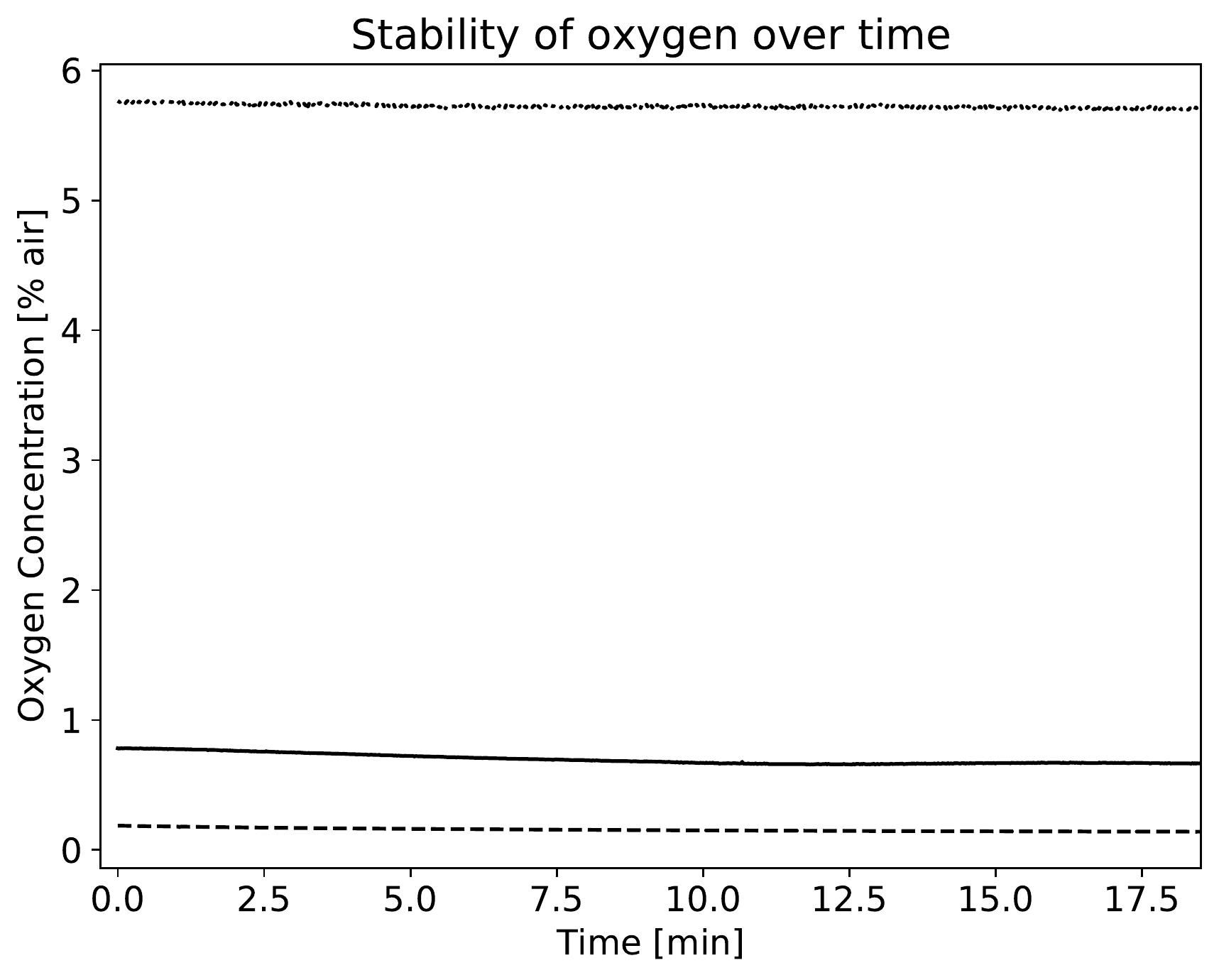}
	\caption{}	\label{fig:control}
	\end{subfigure}
\caption{(a) Oxygen consumption for phantoms with different volumes, irradiated with a dose rate (DR) of 0.0799\,Gy/s. For better visibility, the curves are separated with a time-offset. (b) Oxygen stability was checked in phantom prior to irradiation.}
\end{figure}

At first, different phantoms with 80\,$\mu$l, 583\,$\mu$l, 6.138\,ml and 6.876\,ml water volume were irradiated at a constant dose rate of 0.0799\,Gy/s with 225 kV photons and the concentration of \ch{O2} was measured during irradiation until the concentration of \ch{O2} reached zero. The results of this study are displayed in Figure \ref{fig:volumes}. It was observed that the rate of oxygen removal (i.e. dO/dt) was mostly constant and especially independent on the irradiated volume, as long as the phantoms were irradiated homogeneously. Therefore, due to the observed independence on volume, the phantom's diameter was selected according to the beam's geometries for the following experiments. The stability of \ch{O2} over time in the phantom was tested without radiation for different oxygen levels. Even for low \ch{O2} levels, no diffusion was visible (see Fig. \ref{fig:control}). \\

\subsection{Oxygen Consumption in Photons, Protons and Carbon Ion Beams}
\begin{figure}[h!]
	\centering
	\begin{subfigure}[b]{0.45\textwidth}
		\centering
		\includegraphics[width = \linewidth]{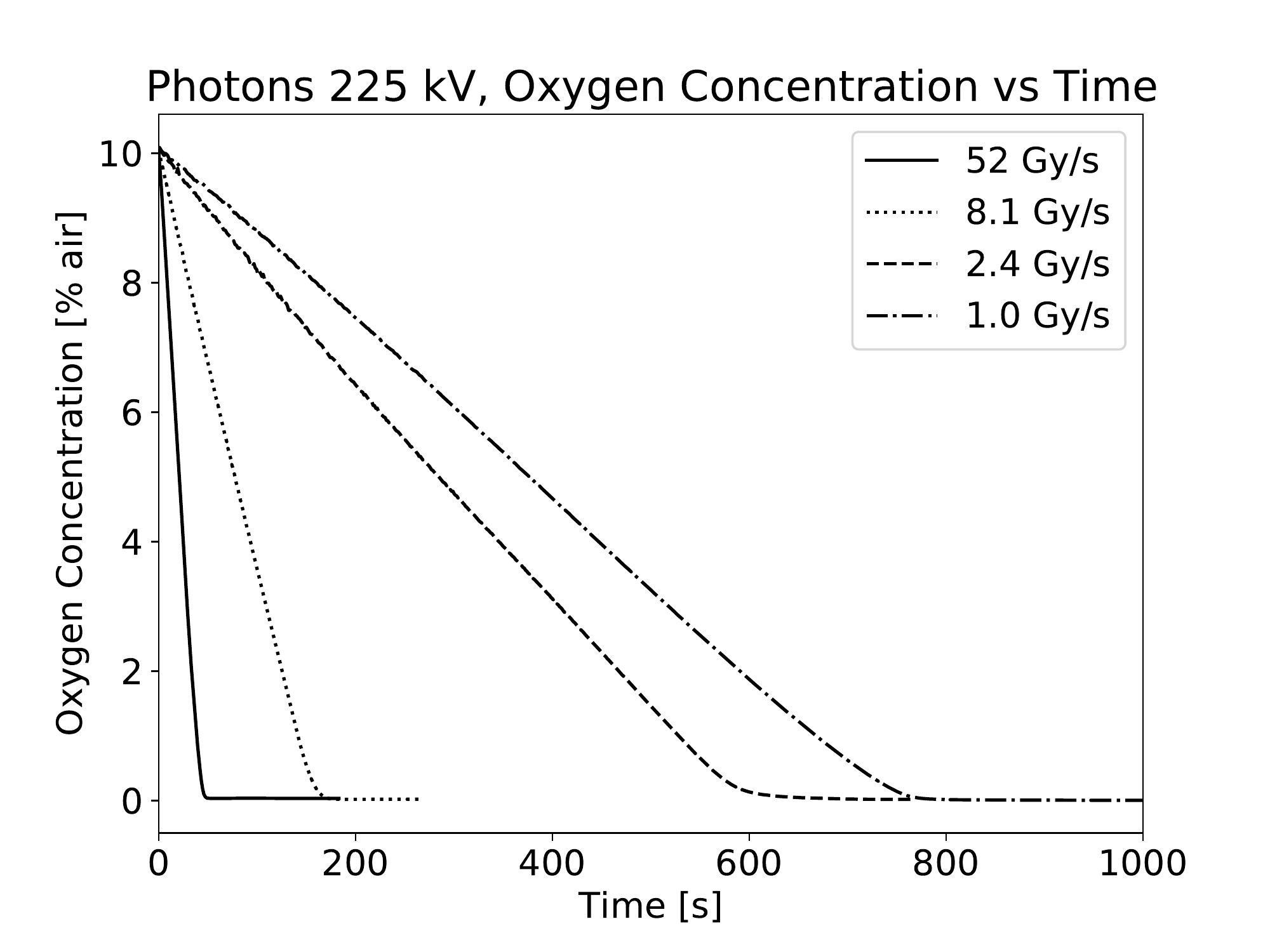} 
\caption{}
\label{fig:doserates_time}
	\end{subfigure}
	\hfill
	\begin{subfigure}[b]{0.45\textwidth}
		\centering
		\includegraphics[width = \linewidth]{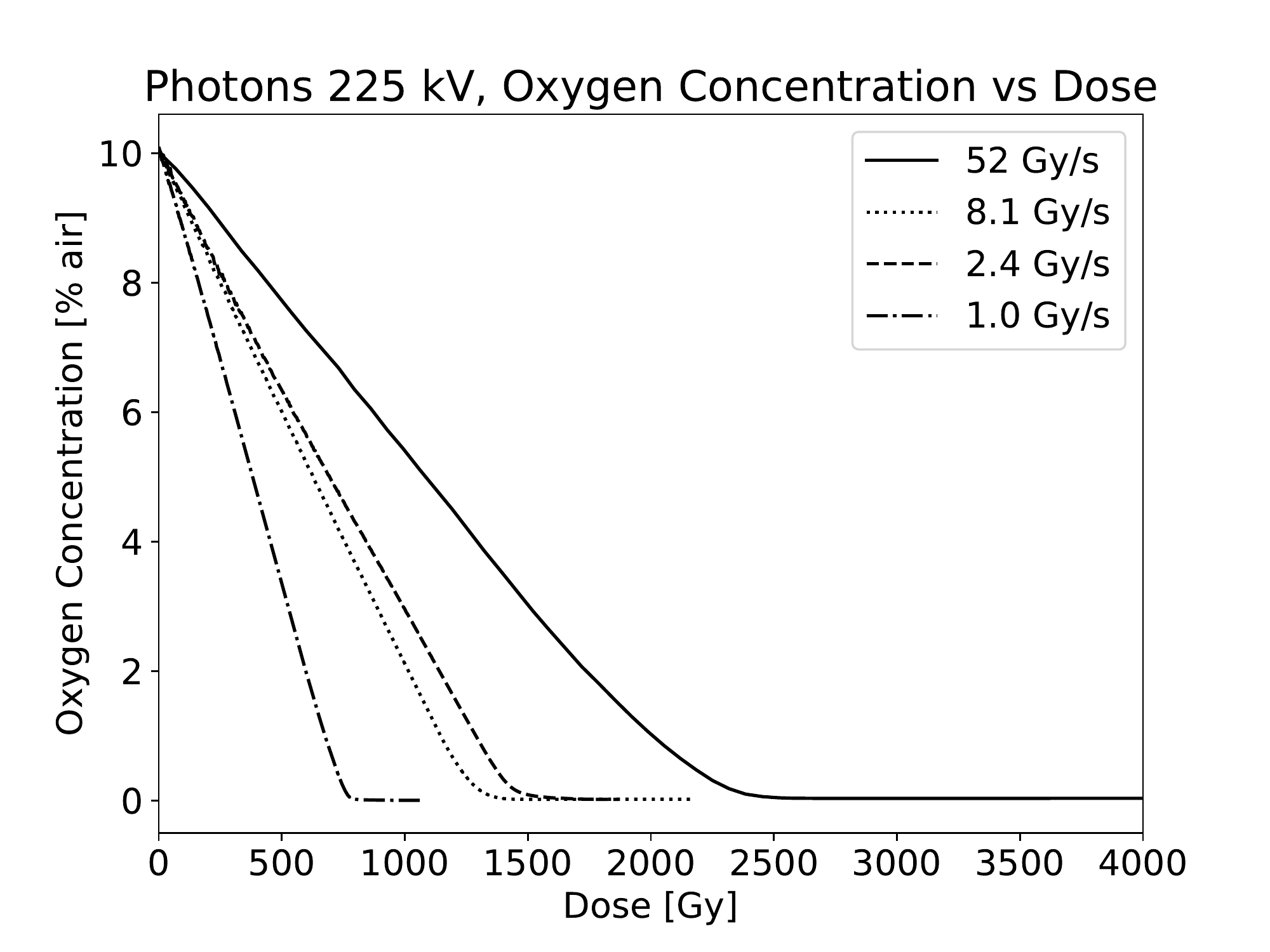} 
\caption{}
\label{fig:doserates_dose}
	\end{subfigure}

	\begin{subfigure}[b]{0.45\textwidth}
	\centering
		\includegraphics[width = \linewidth]{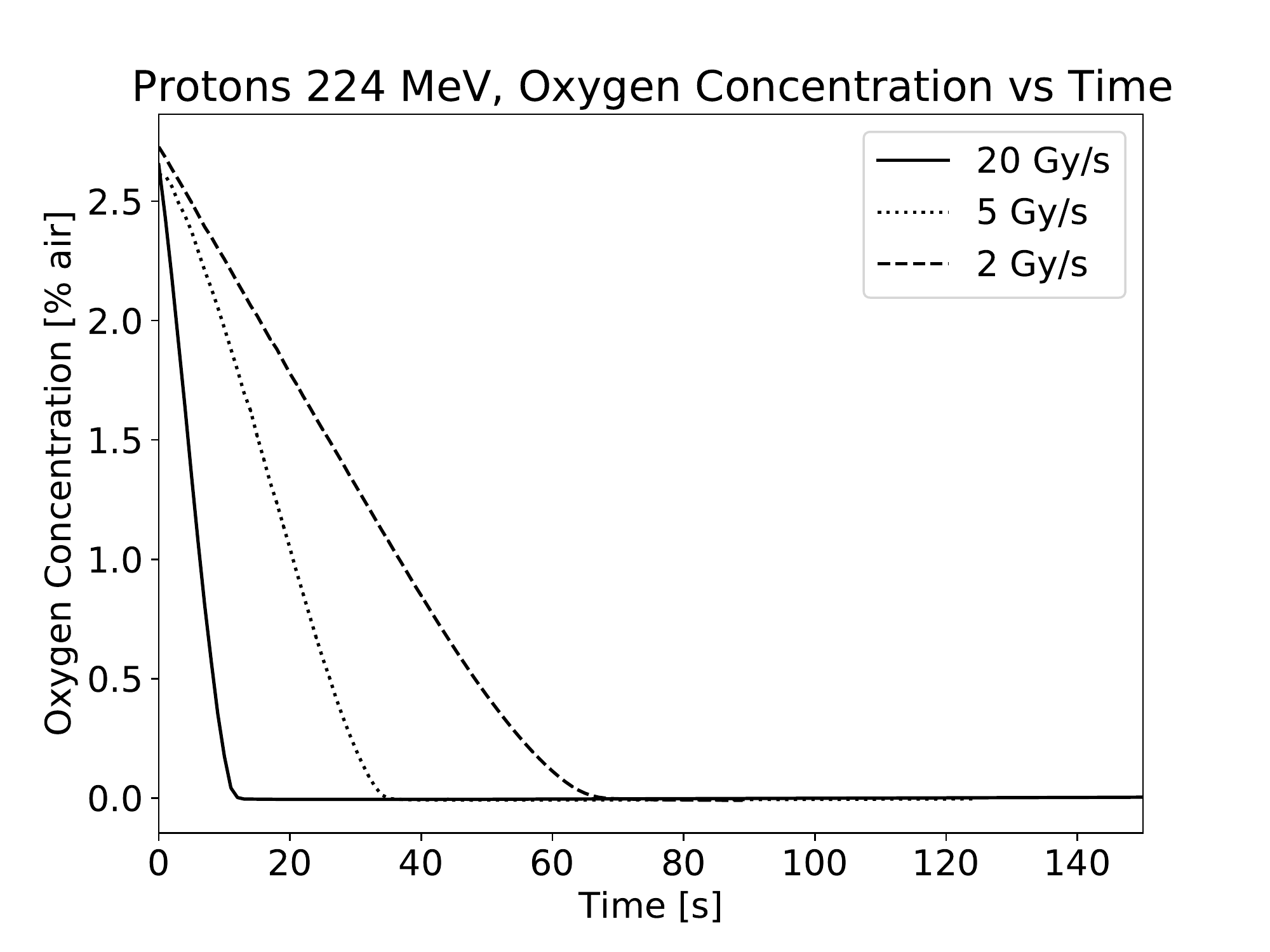} 	
\caption{}
\label{fig:Dresden_time}
\end{subfigure}
\hfill
\begin{subfigure}[b]{0.45\textwidth}
	\centering
		\includegraphics[width = \linewidth]{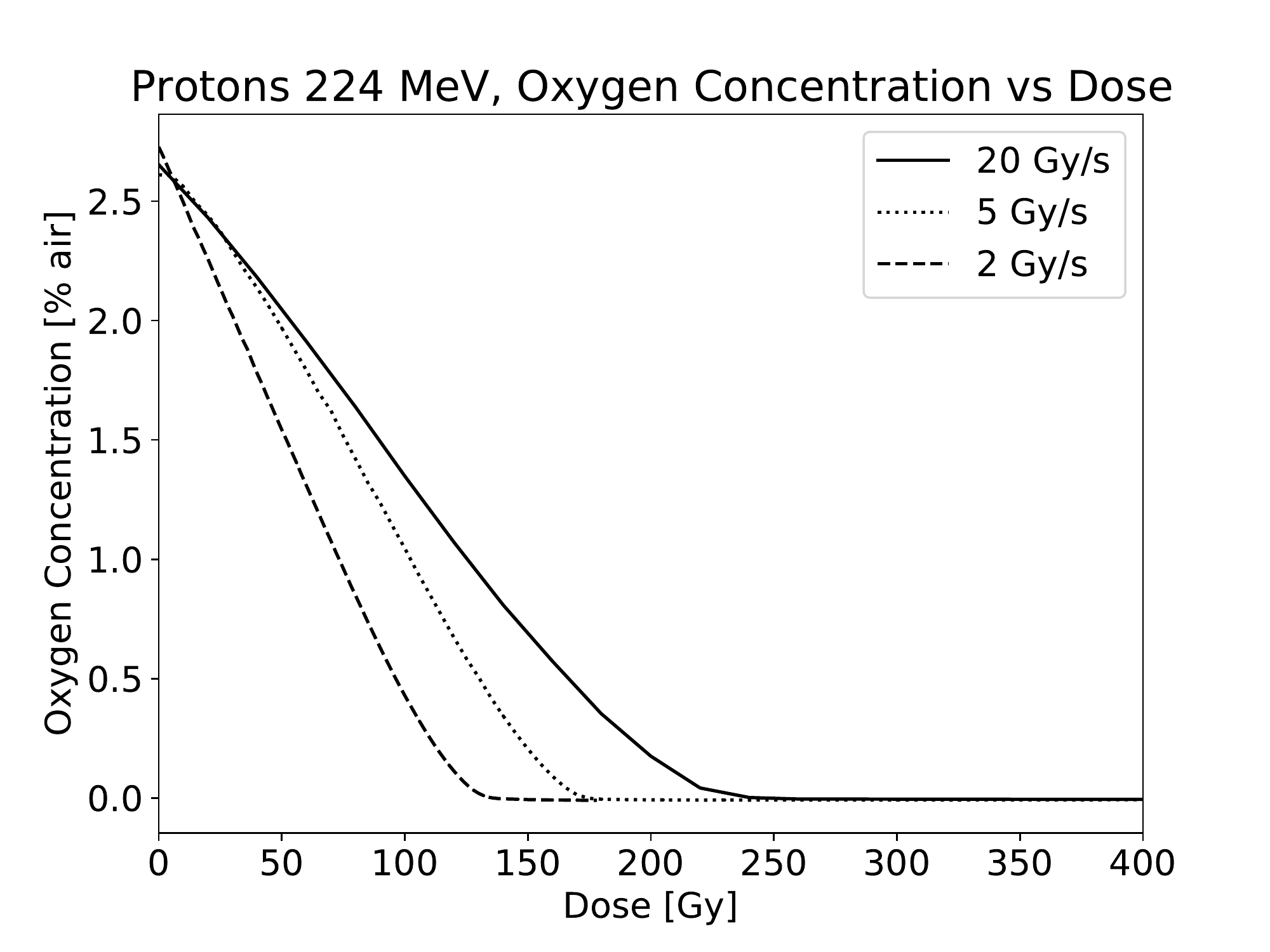} 
\caption{}
\label{fig:Dresden_dose}
\end{subfigure}

	\begin{subfigure}[b]{0.45\textwidth}
	\centering
		\includegraphics[width = \linewidth]{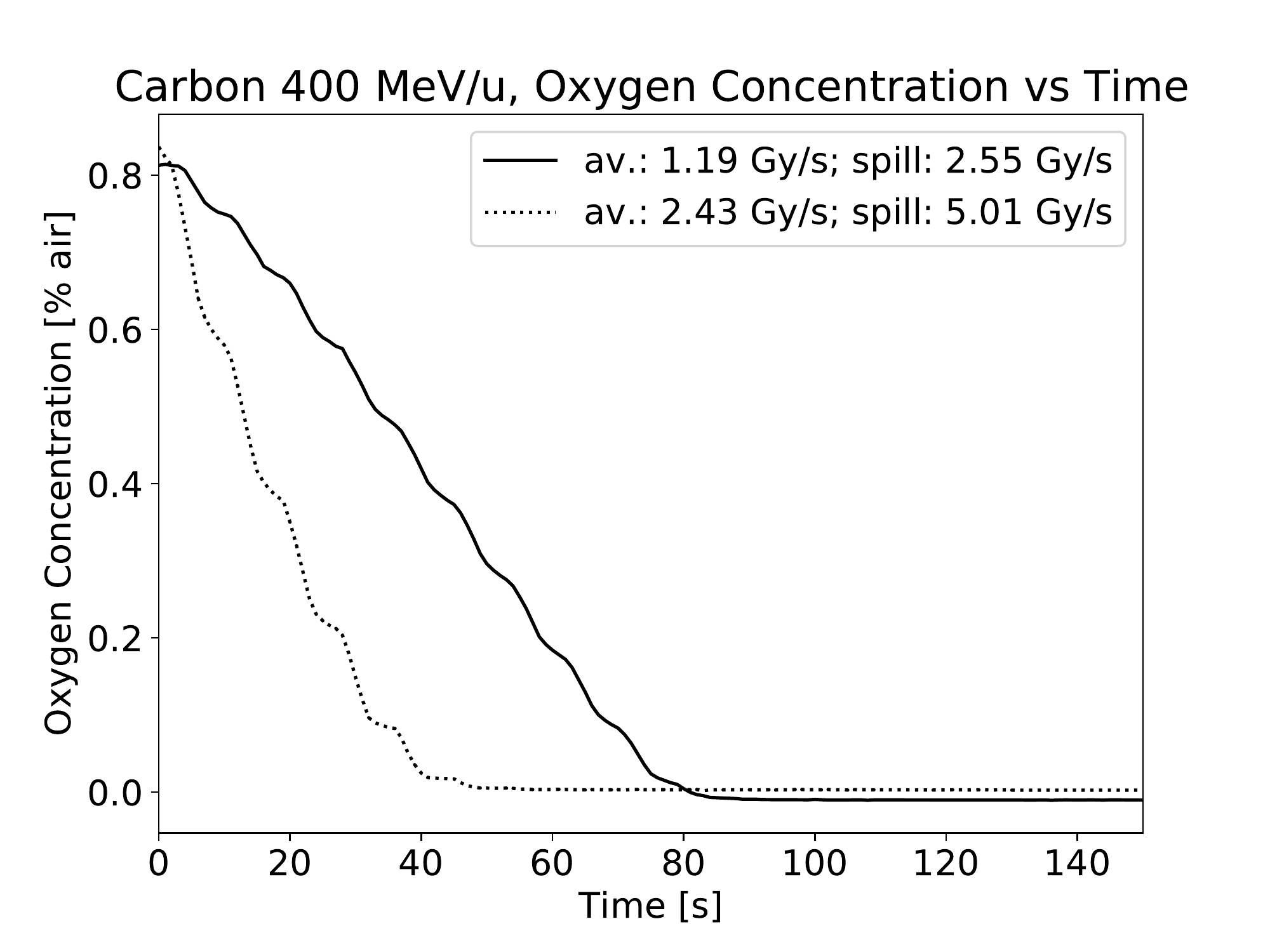} 	
\caption{}
\label{fig:OvsD.png}
\end{subfigure}
\hfill
\begin{subfigure}[b]{0.45\textwidth}
	\centering
		\includegraphics[width = \linewidth]{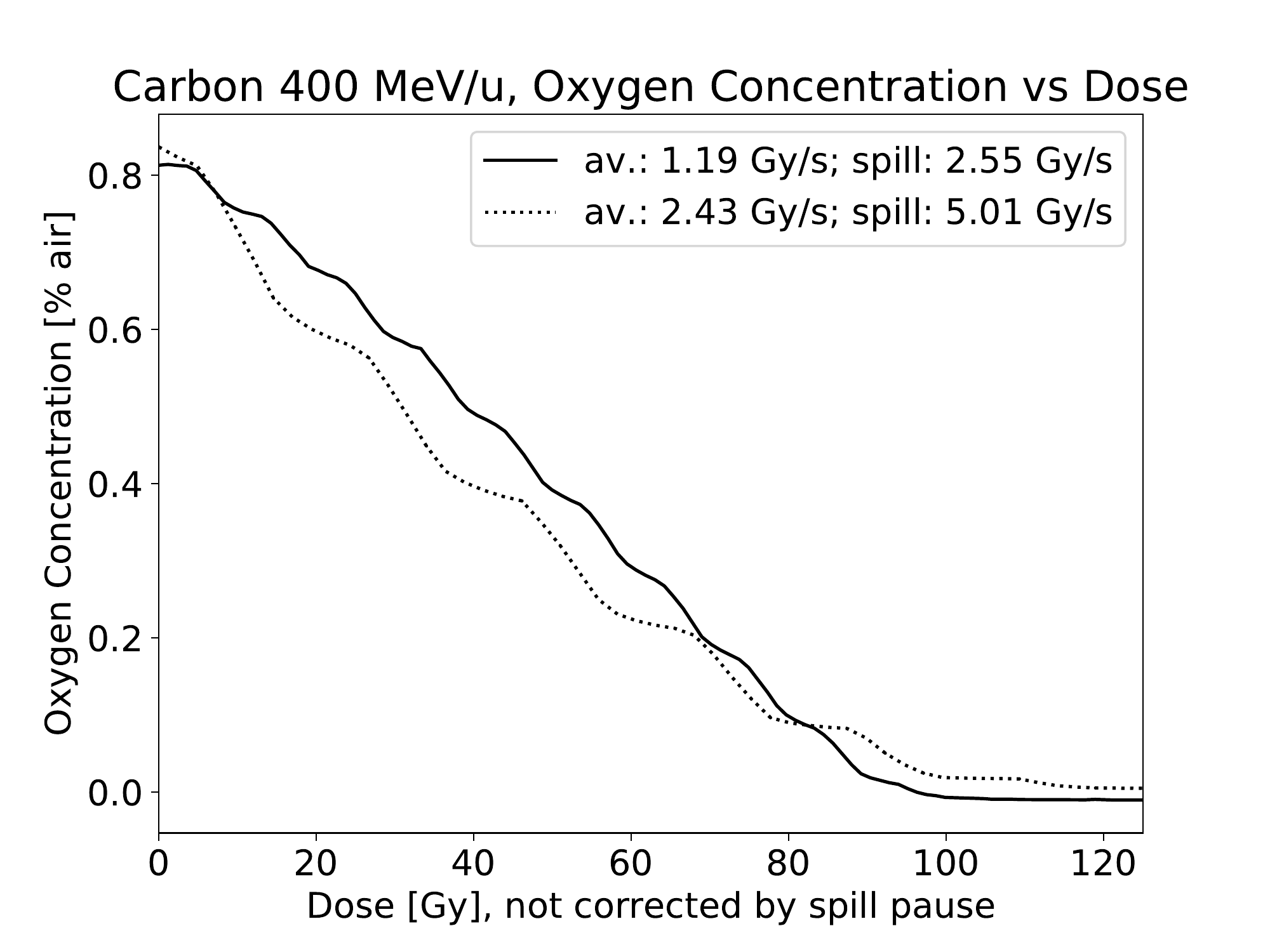} 
\caption{}
\label{fig:OvsDR.png}
\end{subfigure}

	\caption{Oxygen concentration over irradiation time and dose for photon, proton and carbon irradiation. For better comparison, some curves were shifted in time to match the same initial O2 level.}
	\label{fig:doserates_time_dose}
\end{figure}

At all radiation sources, the oxygen consumption in dependence on irradiation time was studied for different dose rates (Fig. \ref{fig:doserates_time}, \ref{fig:Dresden_time}, \ref{fig:OvsD.png}). For photons and protons, the dose dependent oxygen consumption can be achieved by multiplying the time axis with the dose rate present during irradiation (Fig. \ref{fig:doserates_dose}, \ref{fig:Dresden_dose}). 
For carbon ions, beam delivery in spills results in a step-wise oxygen consumption (Fig. \ref{fig:OvsD.png}, \ref{fig:OvsDR.png}). For transferring the time axis into a dose axis, the average dose rate was used. For better comparison, curves were shifted in time to match the same oxygen start level, i.e. irradiation also happened at $t<0$. This was a reasonable simplification as the depletion behavior was mostly linear.

All measurements showed that dose rate had an impact on how fast oxygen got consumed and how much dose was needed for total depletion. Furthermore, the curves showed an almost linear behavior, i.e. the average consumption per unit dose ($\frac{\mathrm{d}O}{\mathrm{d}D}$) was extracted via a linear fit.

\begin{figure}[h!]
	\centering
	\begin{subfigure}[b]{0.45\textwidth}
		\centering
		\includegraphics[width = \linewidth]{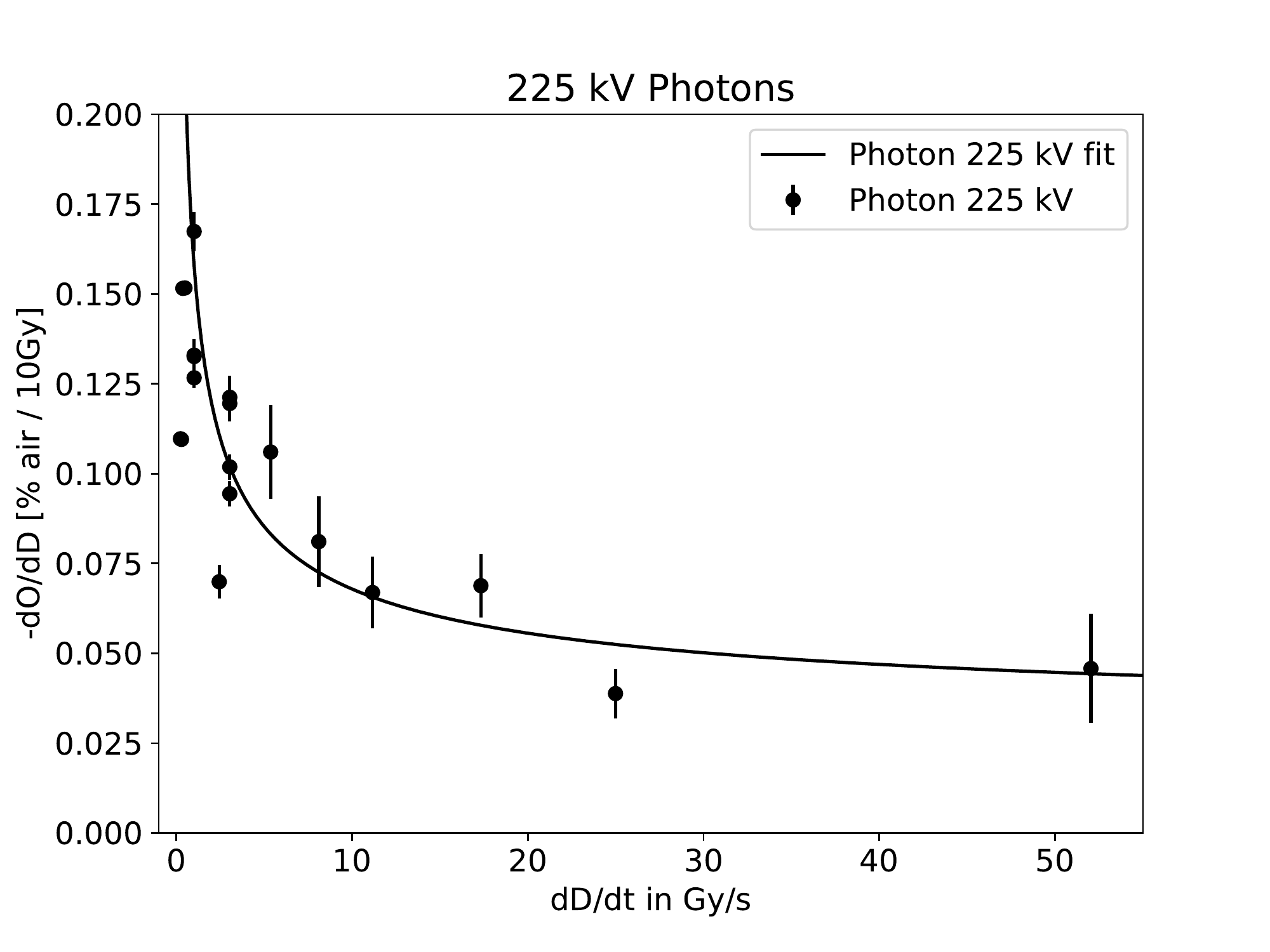} 	
		\caption{}
		\label{fig:photon_dOdt}
	\end{subfigure}
	\hfill
	\begin{subfigure}[b]{0.45\textwidth}
		\centering
		\includegraphics[width = \linewidth]{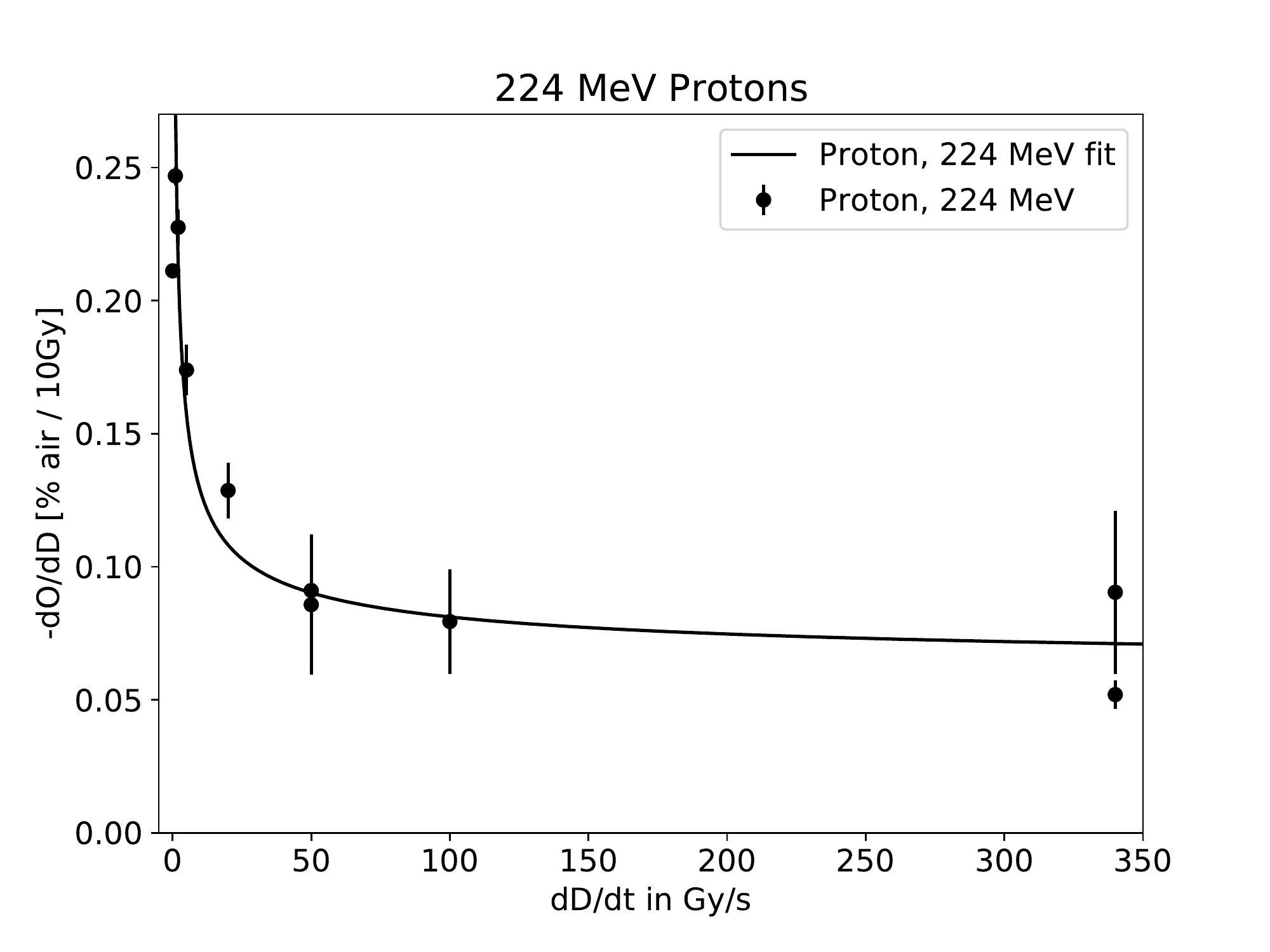} 
		\caption{}
		\label{fig:proton_dOdt}
	\end{subfigure}
\hfill
	\begin{subfigure}[b]{0.45\textwidth}
	\centering
		\includegraphics[width = \linewidth]{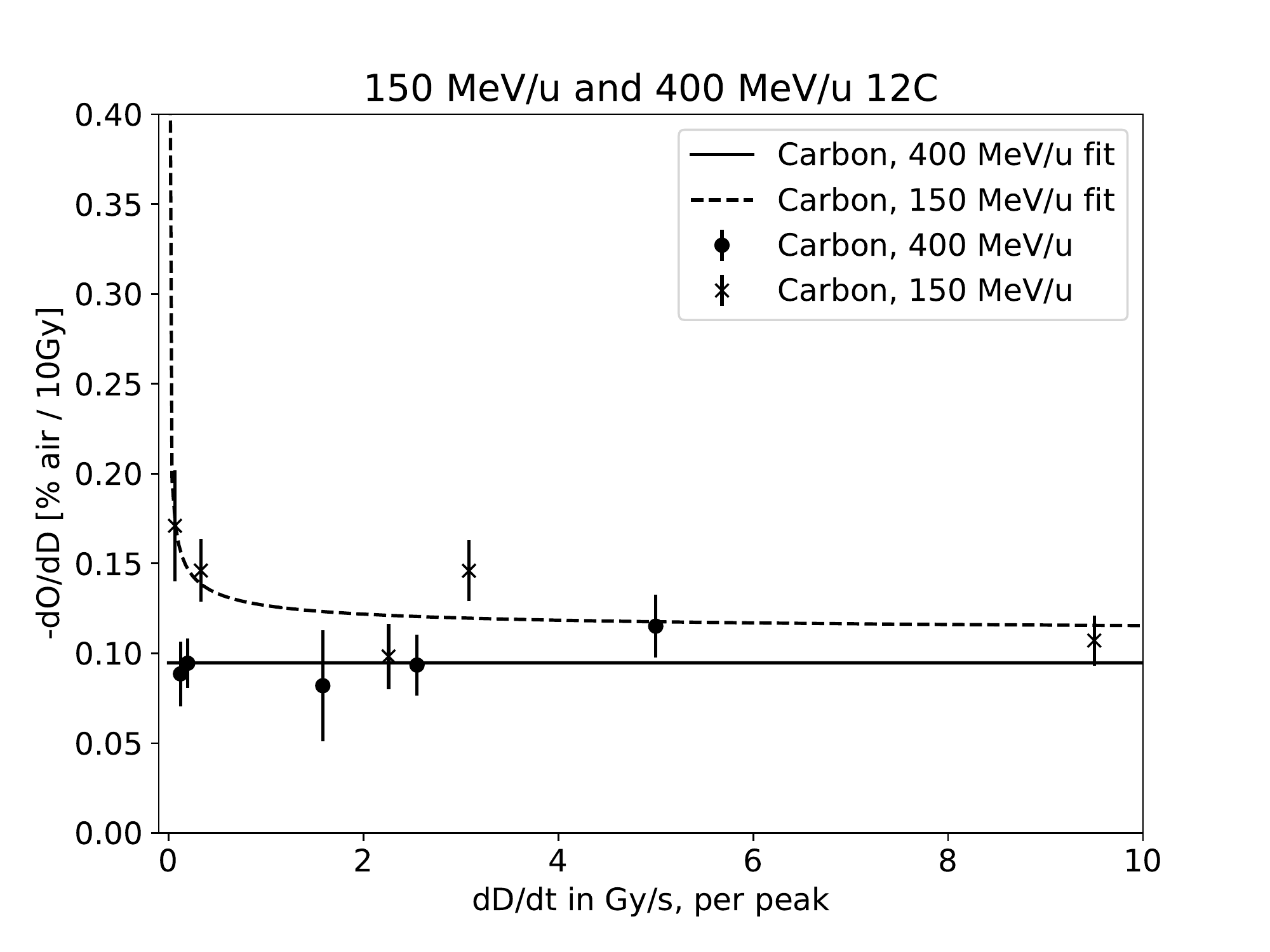} 
	\caption{}
	\label{fig:Carbon_dOdt}
\end{subfigure}
		\caption{Average oxygen consumption per 10 Gy ($\frac{\mathrm{d}O}{\mathrm{d}D}$) as a dependence on dose rate  ($\frac{\mathrm{d}D}{\mathrm{d}t}$) depicted for all measurements in the respective beams types}
	\label{fig:dOdt}
\end{figure}

\clearpage
Subsequently, a broader analysis over multiple measurements was carried out to extract the average amount of depleted oxygen per unit dose ($\frac{\mathrm{d}O}{\mathrm{d}D}$) as a function of dose rate. To obtain the average consumption $\frac{\mathrm{d}O}{\mathrm{d}D}$, the curves from data like exemplarily shown in Figures \ref{fig:doserates_dose}, \ref{fig:Dresden_dose} and \ref{fig:OvsDR.png} were linearly fitted, beginning from the start of irradiation, and the slope fit parameters ($\frac{\mathrm{d}O}{\mathrm{d}D}$) were used for further analysis in Figure \ref{fig:dOdt}. The resulting average consumption $\frac{\mathrm{d}O}{\mathrm{d}D}$ was plotted against the respective dose rate for all radiation types in Figure \ref{fig:dOdt}.  For carbon ion data, $\frac{\mathrm{d}O}{\mathrm{d}D}$ was plotted against the dose rate within one spill and not the average dose rate. 

\clearpage

The fit function used in Figure \ref{fig:dOdt} to describe the amount of consumed oxygen per dose was chosen according to Labarbe et al. \cite{labarbe2020physicochemical} and Mihaljevic et al. \cite{mihaljevic2011linoleic} and is given by a power law, with the parameters $a \leq 0$ and $b>0$:
\begin{equation}
\frac{\mathrm{d}O}{\mathrm{d}D} = a + b\cdot \left(\frac{\mathrm{d}D}{\mathrm{d}t}\right)^{-0.5} \label{Eq:Fit}
\end{equation}

Eq. \ref{Eq:Fit} was used for $^{12}$C data, with $\frac{\mathrm{d}O}{\mathrm{d}D}$ being the average consumption per measurement, derived from average dose rate in the measurement. $\frac{\mathrm{d}D}{\mathrm{d}t}$ described the peak dose rate. For proton and X-ray measurements, peak dose rate and average dose rate were identical, so it was more convenient to work with $\frac{\mathrm{d}O}{\mathrm{d}t}$ data directly from measurement. For this purpose,  Eq. \ref{Eq:Fit} was multiplied by $\frac{\mathrm{d}D}{\mathrm{d}t}$  to obtain Eq. \ref{Eq:Fit_2}. In a second step, fit parameters from Eq. \ref{Eq:Fit_2} were used to generate Fig. \ref{fig:photon_dOdt} and Fig. \ref{fig:proton_dOdt} using Eq. \ref{Eq:Fit}.

\begin{equation}
	\frac{\mathrm{d}O}{\mathrm{d}t} = a \cdot \frac{\mathrm{d}D}{\mathrm{d}t}+ b\cdot \left(\frac{\mathrm{d}D}{\mathrm{d}t}\right)^{+0.5} \label{Eq:Fit_2}
\end{equation}

\begin{figure}[h!]
	\centering
	\begin{subfigure}[b]{0.45\textwidth}
		\centering
		\includegraphics[width = \linewidth]{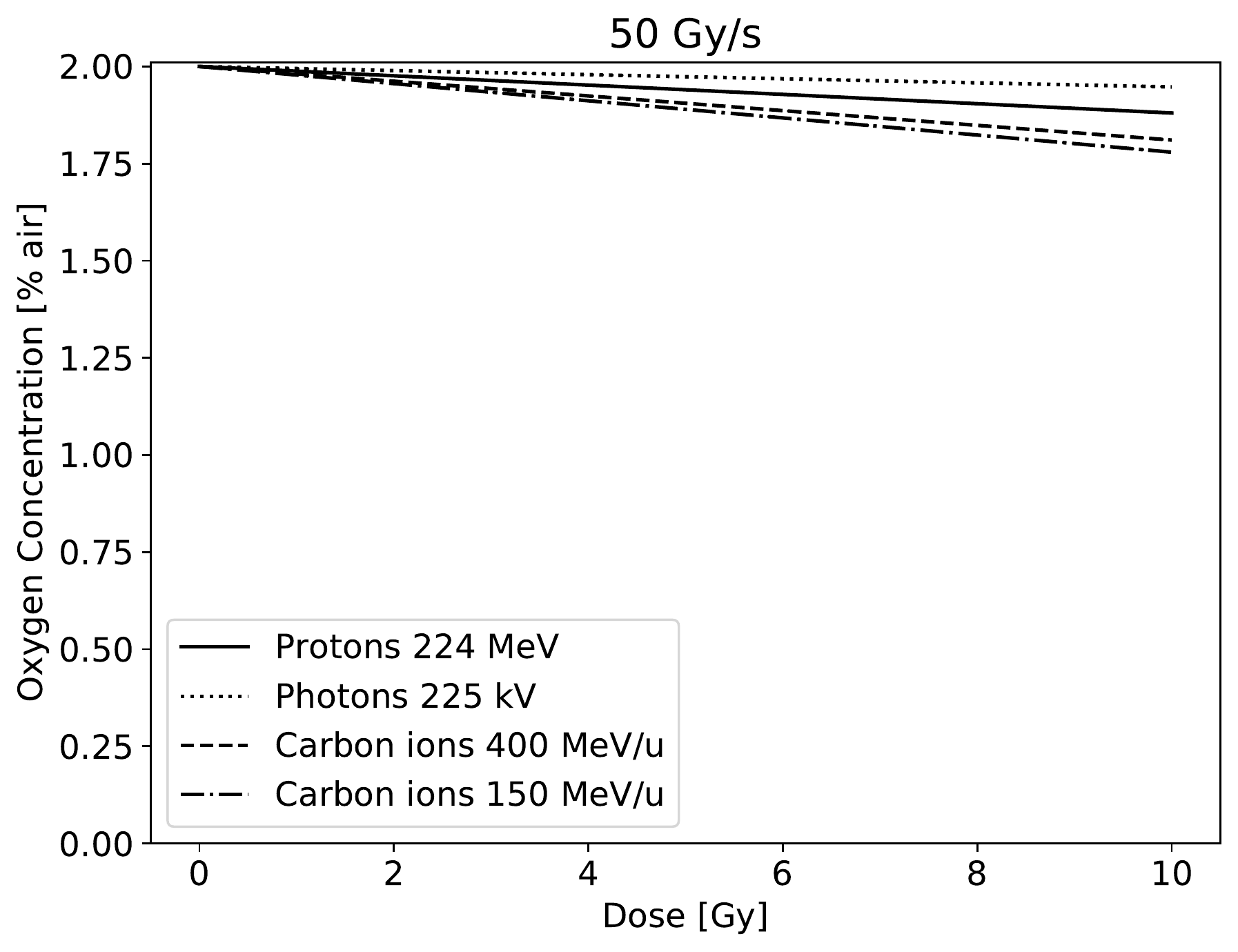}\caption{}\label{fig:fit1}
	\end{subfigure}
	\hfill
	\begin{subfigure}[b]{0.45\textwidth}
		\centering
		\includegraphics[width = \linewidth]{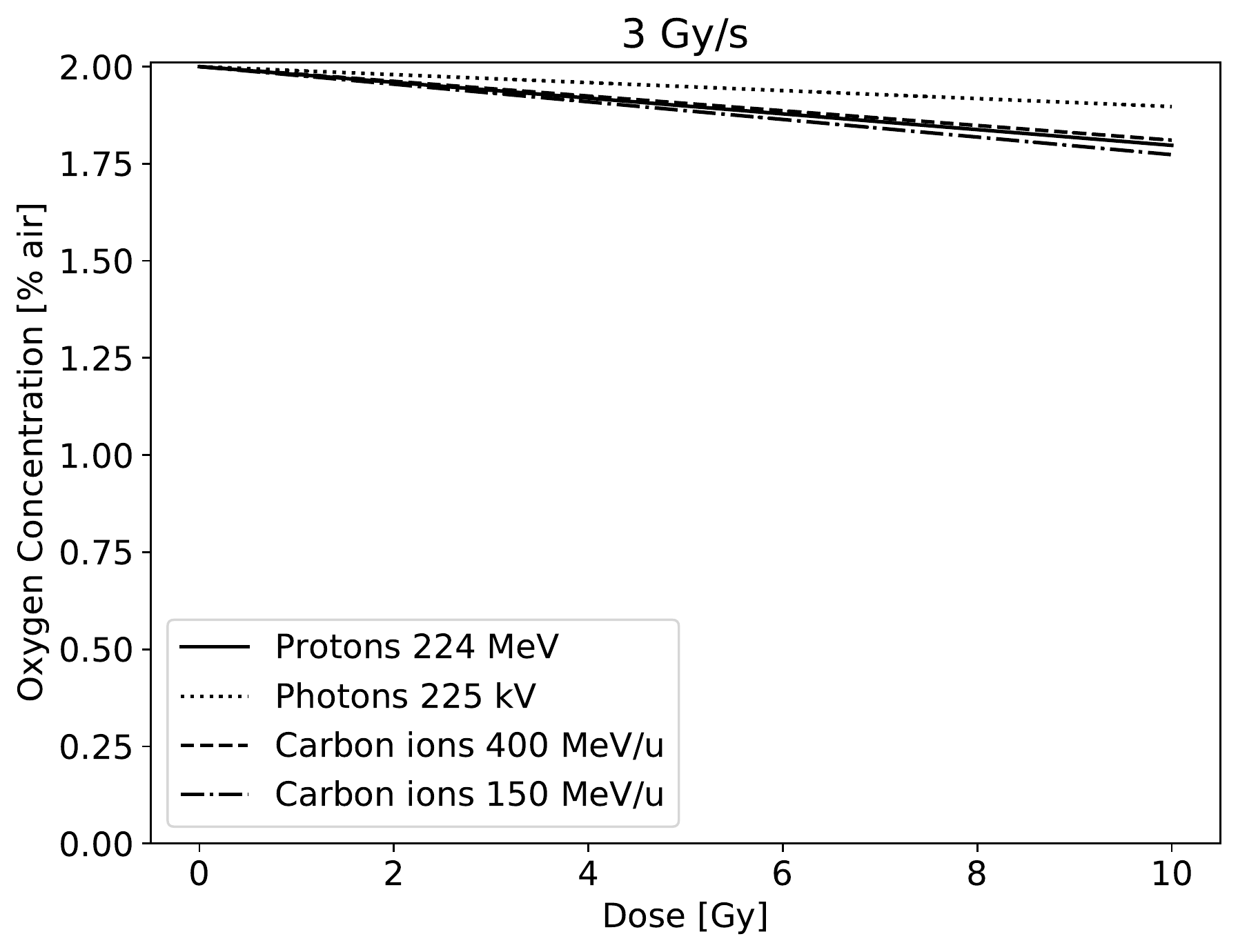}\caption{}\label{fig:fit2}
	\end{subfigure}
	\hfill
	\begin{subfigure}[b]{0.45\textwidth}
		\centering
		\includegraphics[width = \linewidth]{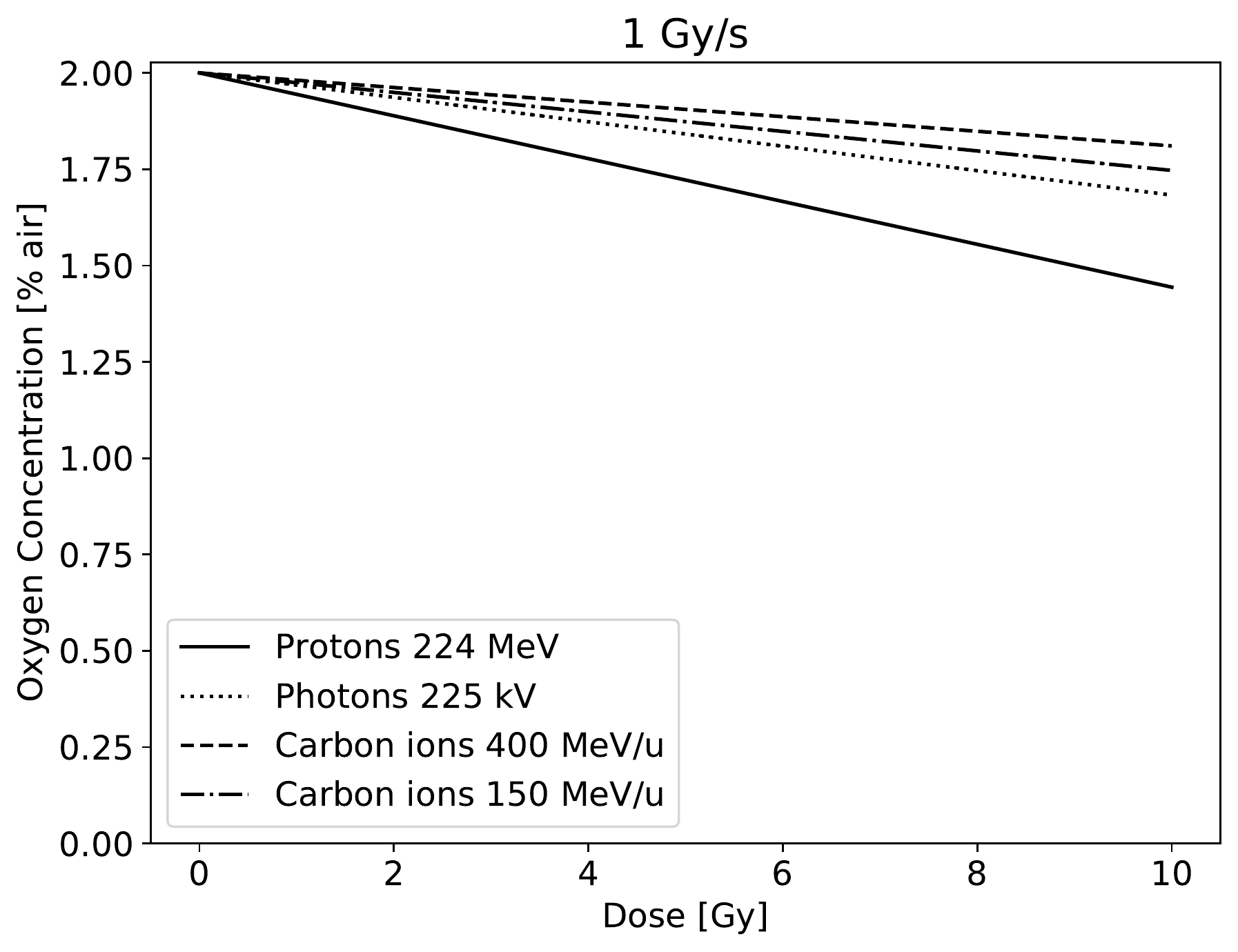}\caption{}\label{fig:fit3}
	\end{subfigure}
	\caption{Linearly extrapolated oxygen consumption starting from 2\,\% atm \ch{O2} for various dose rates. For all settings, large doses (more than 10\,Gy) would be needed to deplete all \ch{O2}.} \label{fig:fit_all}
\end{figure}

  It is evident that all depicted curves in Fig. \ref{fig:dOdt} show a similar curvature, meaning that higher dose rates lead to less oxygen consumption. Furthermore, different beam types have an impact on the oxygen consumption. Figures \ref{fig:fit1} - \ref{fig:fit3} show the fit results per irradiation type applied to a starting oxygen concentration $O_{initial}$ of 2\,\% atm and extrapolated to different dose rates, assuming a linear depletion (which is in reasonable agreement to the measured data). It was used: $	O(D) = O_{initial} - \frac{\mathrm{d}O}{\mathrm{d}D} \cdot D$ with $\frac{\mathrm{d}O}{\mathrm{d}D}$ as parametrized in Equation \ref{Eq:Fit} with fit parameters $a$ and $b$ obtained from fits shown in Figure \ref{fig:dOdt}.
  
  Considering Figure \ref{fig:fit_all}, it is evident that 2\,\% atm \ch{O2} cannot be depleted within 10\,Gy by any of the used beams.

\section{Discussion}

The aim of the study was to investigate if oxygen depletion occurs during (FLASH-) irradiation, by measuring the oxygen concentration in-vitro during irradiation of water by photons, protons and carbon ions. For all experiments, TROXSP5 sensors were used to measure the oxygen concentration during irradiation, with the sensors not being affected by the radiation.\\
The oxygen consumed during irradiation was found to be linear in time and dose and independent of the initial oxygen concentration (Fig. \ref{fig:doserates_time}-\ref{fig:OvsDR.png}). Oxygen consumption was evaluated per total dose delivered, where we defined  $\frac{\mathrm{d}O}{\mathrm{d}D}$ to represent the total oxygen removed per unit dose (Fig. \ref{fig:doserates_dose}, \ref{fig:Dresden_dose}, \ref{fig:OvsDR.png}). The  $\frac{\mathrm{d}O}{\mathrm{d}D}$ represents also the total amount of radicals produced per total dose delivered, which have reacted with the diluted oxygen to create a reactive oxygen species (ROS). The  $\frac{\mathrm{d}O}{\mathrm{d}D}$ as a function of dose rate (Fig. \ref{fig:photon_dOdt}-\ref{fig:Carbon_dOdt}) follows the same non-linear behavior as described by Labarbe et al.\cite{labarbe2020physicochemical} in organic environments, and given by the expression $a+b \cdot \left(\frac{\mathrm{d}D}{\mathrm{d}t}\right)^{-0.5}$. The nonlinearity was described in Labarbe et al. as the self-interaction of \ch{ROO^.} molecules, part of the termination reaction. In the present study, experiments were carried out with only pure water and therefore without the presence of any \ch{RH} or \ch{R^.} molecules. However, in pure water, self-interactions of the radicals play a major role in removing the radicals that could potentially react with oxygen. This applies especially to \ch{e^-_{aq}} and \ch{H^.} as described in the following:

\begin{equation}
\ch{e^-_{aq} + e^-_{aq} + 2 H2O -> H2 + 2 OH-}
\end{equation}
\begin{equation}
\ch{e_{aq}^- + O2 -> O2^{.-}} 
\end{equation}
\begin{equation}
\ch{H^. + H^. -> H2}\newline
\end{equation}
\begin{equation}
\ch{H^. + O2 -> HO_2^.} 
\end{equation}

The presence or absence of \ch{e^-_{aq}} yields directly a change in \ch{O2} as described in the reaction (4). Therefore one would expect that an increased dose rate should yield a higher \ch{O2} consumption because of the higher production of \ch{e^-_{aq}} per second. However, because of the competing reaction (3),  \ch{e^-_{aq}} is removed faster with increased dose rate yielding a lower steady state of \ch{e^-_{aq}}.  The lower steady state of \ch{e^-_{aq}} means there is less of \ch{e^-_{aq}} available to react with \ch{O2} via reaction (4). The same process applies to \ch{H^.}, as described in reactions (5) and (6).  

The amount of radicals produced is given by $G\cdot C \cdot \frac{\mathrm{d}D}{\mathrm{d}t}\cdot t$ with $G$ being the $G$-value, $C$ is a constant ($1.04\times10^{-7} \mathrm{\frac{mol}{dm^3}s} \times \left[G\right]^{-1}\mathrm{(\frac{Gy}{s})^{-1}}$)\cite{joseph2008combined}, $\frac{\mathrm{d}D}{\mathrm{d}t}$ is the dose rate and $t$ the irradiation time. In order to understand if radicals can self-interact we need to check the mean free path length of the radicals. The radicals will diffuse for an average path length $R_{rms}$ of $2.26 \cdot \sqrt{D_{\mathrm{diff}} \cdot t_{\frac{1}{2}}}$(\cite{roots1975estimation}) with $t_{\frac{1}{2}}$ being the half-life and $D_{\mathrm{diff}}$  being the diffusion constant. Typical values for \ch{e_{aq}^-} are  $D_{\mathrm{diff}} = 4.5\times 10^{-9} \mathrm{\frac{m}{s^2}}$, and $t_{\frac{1}{2}} = 47\,\mu \mathrm{s}$ (\cite{roots1975estimation}, \cite{Dal_Bello}, \cite{walker1967}). From these values, a mean free path length of $\sim 1\,\mu \mathrm{m}$ can be estimated. Hence, it becomes clear that the solvated electrons \ch{e_{aq}^-} diffuse far enough to interact with each other, ultimately reducing the steady state's concentration. 
\textbf{Therefore, \ch{O2} consumption is reduced with increasing dose rate as observed in our experiments.}

The study presented here showed that for FLASH dose rates, radical recombinations, via reaction (3) and (5), reduce the oxygen consumption within the medium. In addition, we observed experimentally that the amount of oxygen consumed by radiation depends also on the particle type and its LET but further investigation would be needed.
For the case of 10\,Gy dose delivery, the amount of oxygen consumed was 0.04\,\% atm - 0.18\,\% atm for 225\,kV photons, 0.04\,\% atm - 0.25\,\% atm for 224 MeV protons and 0.09\,\% atm - 0.17\,\% atm for carbon ions, dependent on the dose rate (Fig. \ref{fig:photon_dOdt}-\ref{fig:Carbon_dOdt}). The obtained experimental values are in good agreement with other published results of experiments in water, where oxygen consumption was between 0.26\,\% atm to  0.42\,\% atm \cite{Weiss1974, evans1969, day1949chemical, Whillans, michaels1986oxygen} for photon radiation. Recent modeling studies also yielded oxygen consumption between 0.05\,\% atm up to 0.27\,\% atm \cite{labarbe2020physicochemical, boscolo2020impact, pratx2019computational} for photon, proton and carbon ion beams.  In addition, a theoretical prediction by Petersson et al. \citep{petersson2020quantitative} yields an oxygen consumption in the range of 0.1\,\% atm to 2\,\% atm for total delivered dose of 10\,Gy with FLASH. Applying the experimental findings and curves of Fig. \ref{fig:photon_dOdt}-\ref{fig:Carbon_dOdt} onto an exemplary case of a water phantom with 2\,\% atm \ch{O2} content, it is evident that 10\,Gy radiation of any analyzed radiation type cannot deplete oxygen completely in water (Fig. \ref{fig:fit1}-\ref{fig:fit3}). It can be concluded, that for higher FLASH dose rates, less oxygen depletion per dose was observed.

\section{Conclusion}
This study showed that TROXSP5 sensors are a suitable sensor type to measure oxygen consumption during radiation non-invasively in water phantoms. No total depletion of oxygen was observed for 10\,Gy delivery by FLASH irradiation for photons, protons and carbon ions. Hence, oxygen depletion is not a suitable mechanism to explain the FLASH effect alone but rather a reduction of oxygen consumption was found for higher dose rates which was related to the lower steady state values of \ch{e^-_{aq}} radicals. The results presented here are in good agreement with previous data and recent radio-chemical models but the outcome stresses \textbf{non-linear dose rate dependence} of the oxygen consumption, even without the presence of organic molecules, which is to date not implemented in current models. 

\section{Acknowledgments}
This work was supported by the Deutsche Forschungsgemeinschaft (GSC129). Furthermore, this project has received funding from the European Union's Horizon 2020 research and innovation program under grant agreement No. 730983 (INSPIRE). \\
This work was also supported by grants of the German-Israeli Helmholtz Research School in Cancer Biology – Cancer Transitional and Research Exchange Program (Cancer-TRAX).\\
The authors would like to thank Dr. Peter Häring and Mona Lifferth from the department of Physical Quality Assurance in Radiation Therapy, DKFZ for help with the dosimetry.

\section{Conflict of Interest}
The authors declare no conflict of interest.

\newpage     

\clearpage

\section*{Appendix}

\clearpage
\section*{References}
\addcontentsline{toc}{section}{\numberline{}References}
\vspace*{-20mm}


\bibliographystyle{ama}    

\listoffigures

\end{document}